\documentclass{ckm}                 

\confname{Workshop on the CKM Unitarity Triangle, IPPP Durham, April 2003}

\usepackage{epsfig}

\input{babarsym}

\def\BR         {{\ensuremath{\cal B}\xspace}}
\def\B       {\ensuremath{B}\xspace}
\def\Bzb     {\ensuremath{\Bbar^0}\xspace}
\def\Bbar    {\kern 0.18em\overline{\kern -0.18em B}{}\xspace}

\def\epem       {\ensuremath{e^+e^-}\xspace}

\def\jpsi     {\ensuremath{{J\mskip -3mu/\mskip -2mu\psi\mskip 2mu}}\xspace}

\def\Vcb  {\ensuremath{|V_{cb}|}\xspace}
\def\Vub  {\ensuremath{|V_{ub}|}\xspace}

\title{A Preliminary Measurement of Hadronic Mass Moments in Semileptonic \B Meson Decays}

\author{Vera G. L\"{u}th\addressmark{a} 
\thanks{Work supported by the US Department of Energy}
representing the \babar\ Collaboration
}

\address[a]{Stanford Linear Accelerator Center\\
PO Box 20450, Stanford, CA 94309, USA}

\begin{document}

{\pagestyle{empty}


\begin{abstract}
A preliminary measurement of the second
moments of the hadron mass in $B \to X \ell \nu$ decays by the \babar\
Collaborations is reported.  These measurements are performed as a function of the lepton momentum above a given threshold.
\vspace{1pc}
\end{abstract}

\maketitle

For many years semileptonic decays have been the topic of a large variety of
studies because they are theoretically simple at the parton level,
sensitive to the coupling of quarks to the weak charged current,
and allow us to probe the impact of strong interactions on the
bound quark. Experimentally they are readily accessible, because
of the large branching fraction and clear signature in form of a
high momentum lepton. The principal goal for studies of semileptonic \B\ meson decays is the
determination of the CKM matrix elements \Vcb and \Vub.

The decay width for inclusive semileptonic \B\ decays to the charm
states $X_c$ can be written as 
\begin{equation}
\Gamma^c_{SL} \equiv \BR(\overline{B} \ra X_c \ell^-
\overline{\nu})/\tau_B
               = \gamma_c |V_{cb}|^2,
\end{equation}
i.e., $|V_{cb}|$ can be determined from the branching fraction and
the average \B\ lifetime, provided the factor $\gamma_c$ is known. The
theoretical QCD parameter $\gamma_c$ requires both perturbative
and non-perturbative input. In the framework of Heavy Quark
Effective Theory the uncertainties in the estimate of $\gamma_c$
can be reduced by using information from other inclusive measurements,
for instance, the moments of the mass distribution of the hadrons $X_c$. Like the total decay rate, this inclusive observable
can be calculated using expansions in powers of the strong
coupling constant $\alpha_s(m_b)$ and in inverse powers of the $B$
meson mass, $m_B$, that include non-perturbative parameters.  At
order $1/m_B^2$, there are three parameters, $\bar{\Lambda}$,
$\lambda_1$, and $\lambda_2$. From the $B^*-B$ mass splitting, we
have $\lambda_2=0.128\pm 0.010\gev^2$. 

In the following, we report a preliminary measurement performed with the
\babar\ detector \cite{BABARDET} operating at the \FourS\
resonance at the \pep2\ energy asymmetric \epem storage ring
\cite{PEPII} at SLAC.  This measurement of the second moment of the hadron mass
distribution as a function of the minimum lepton momentum was first reported last summer~\cite{ichep2}. An update of this measurement is expected for this summer's conferences.



We measure the second moment of the invariant mass $M_{X}$
distribution of the hadronic system $X$ in $B \to X \ell \nu$  decays, $\langle M_X^2
- \overline{m}_D^2 \rangle$, where $\overline{m}_D = (m_D + 3
m_{D^*})/4 = 1.975~\gevcc$ is the spin-averaged $D$ meson mass.
This measurement is similar to one performed by CLEO~\cite{Cronin-Hennessy}.

The analysis is based on a sample of 55 million \BB\ events, from
which we select a subsample of 5,800 events (above a background of 3,600 events that are statistically subtracted using the energy constrained \B\ mass distribution).  In these events one $B$ meson is fully reconstructed in a
hadronic decay mode and the semileptonic decay of the second \B\
is identified by a high momentum electron or muon. The 
system $X$ in the decay $B \ra X \ell \nu$ is made up of hadrons and
photons that are not associated with the $B_{reco}$ candidate. We
exploit the available kinematic information of the full \BB\ event by performing a 2C kinematic fit
that imposes four-momentum conservation, the equality of the
masses of the two $B$ mesons,
and forces $M_{miss}^2 = M_{\nu}^2 = 0$. The fit takes into
account event-by-event the measurement errors of all individual
particles and the measured missing mass. This leads not only to a
significant improvement of the r.m.s. of the mass resolution of
the $X$ system but also provides an almost unbiased estimator of
the mean $M_X$ and a resolution that is largely independent of
$M_{miss}^2$.\\

Figure~\ref{fig:mxdist} shows the resultant $M_X$ distribution of
the selected events, for a minimum lepton momenta $P^*_{min}=0.9
\gevc$. Different $B \rightarrow X_c l\nu$ decays contribute to
this distribution. The dominant decays are $B \rightarrow D^*
l\nu$ and  $B \rightarrow D l\nu$, but we also expect
contributions from decays to higher mass charm states, $D^{**}$
resonances with a mass distribution $X_H^{reso}$ peaked near 2.4
\gevcc, and potentially non-resonant $D^* \pi$ final states for
which we assume a broad mass distribution $X_H^{nreso}$ extending
to the kinematic limit. The background is dominated by secondary
semileptonic charm decays, contributions from lepton (primarily
muon) misidentification are much smaller. The backgrounds decrease
significantly for higher $P^*_{min}$.

\unitlength1.0cm 
\begin{figure}[h]
\mbox{\epsfig{file=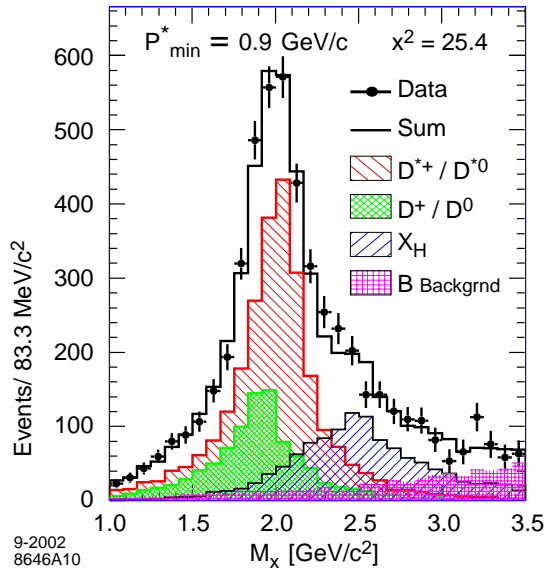, scale=.9}} 
\vspace{-0.5cm}
\caption{The measured $M_X$ distribution for $P^*_{min}=0.9
\gevc$. The hatched histograms show the fitted contributions from
$B \rightarrow D^* l \nu$, $B \rightarrow D l \nu$ and $B
\rightarrow X_H l \nu$ decays, as well as the background
distribution. The white histogram represents the sum of all the
individual distributions.}
\label{fig:mxdist}
\end{figure}

A binned $\chi^2$-fit to the $M_X$ distribution is performed to
determine the relative size of the three signal contributions,
$f_{D*}$, $f_D$, and $f_{X_H}$, where $f_{X_H}$ refers to the sum
of the resonant and non-resonant high mass charm states.
Taking into account the true particle masses (the $D$ and $D^*$
masses are basically $\delta$ functions, and the mean of the $X_H$
contribution is taken from generated events) the second moments are
calculated according to the following expression
\begin{eqnarray}
\lefteqn \langle M_X^2 - \overline{m}_D^2 \rangle & = & f_{D^*}
\cdot (M_{D^*}^2-\overline{m}_D^2) \nonumber\\ &  & +
f_{D}\cdot(M_{D}^2-\overline{m}_D^2)\nonumber\\ & & +f_{X_H} \cdot
\langle M_{X_H}^2-\overline{m}_D^2 \rangle.
\end{eqnarray}

Figure~\ref{fig:momentsivar2} shows the second moment as a function
of the lepton momentum above a minimum $P^*_{min}$. The increase at lower
momenta is attributed to contributions from the high mass states,
i.e. the non-resonant $X_H^{nreso}$ decays. The CLEO Collaboration
has measured the same moment for $P^*_{min} =1.5~\gevc$, based on similar assumption about the high mass hadronic states.  Their
result~\cite{Cronin-Hennessy} is in good agreement with the measurement
presented here.  The data are also consistent with a measurement by the
DELPHI Collaboration~\cite{calvi} that corresponds to $P^*_{min}=0 GeV/c$. \\

\begin{figure}[h]
\mbox{\epsfig{file=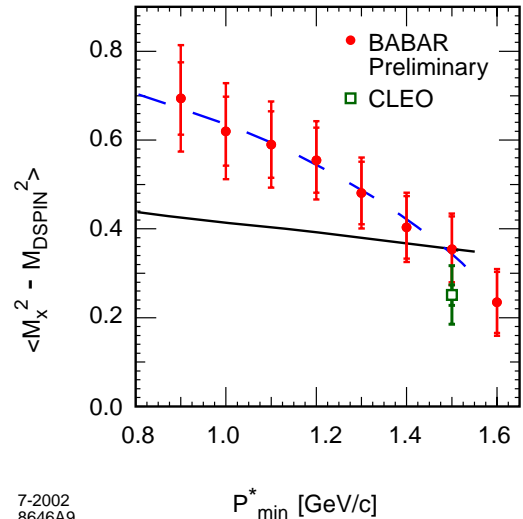, scale=.9}}
\vspace{-.5cm} 
\caption{Measured mass moments as a function of the
minimum lepton momentum, $P^*_{min}$. The errors indicate the sum
of the statistical and systematic uncertainties, they are highly
correlated.
The dashed curve shows the best description of the data by the OPE expansion
\cite{Falk:1998jq} with $\lambda_1$ and $\bar{\Lambda}$ as free
parameters. For comparison, the solid line shows variation of
the moments for the parameters $\lambda_1 = -0.17~\gev^2$ and
$\bar{\Lambda} = 0.35~\gev$ \cite{Chen:2001fj}.}
\label{fig:momentsivar2}
\end{figure}

Extensive studies have been performed to assess the systematic
uncertainties and potential biases in the moment measurement.
These studies involve both changes in the event selection and
variations of the corrections for efficiencies and resolution, and
comparisons of data with Monte Carlo simulations. The leading
systematic error is due to the uncertainty in the model for the
higher mass states $X_H$. Other  errors are due to
uncertainties in the Monte Carlo simulation of the detector
resolution and efficiencies as well as in the background
contributions.

As one of the many cross checks we use the relative contributions
$f_i$ to determine branching fractions by correcting for
acceptance and setting the total semi-leptonic branching fraction
to 10.87\%. The resultant partial branching fractions are fully
compatible with previous measurements and independent of
$P^*_{min}$. This is also the case when we split the $X_H$
contribution into a resonant and non-resonant part and allow both
of them to vary independently.

Heavy Quark Effective Theory (HQET) calculations of the second mass
moment $\langle M_X^2 - \overline{m}_D^2 \rangle$ have been
carried out \cite{Falk:1998jq} using Operator Product Expansions
(OPE) up to order $\alpha^2_s \beta_0$ and $1/m_B^3$. These
expansions contain the non-perturbative parameters $\bar{\Lambda}$
(${\cal O}(m_B)$), $\lambda_1$ and $\lambda_2$ (${\cal
O}(m^2_B)$). The observed dependence of the moments on the minimum
lepton momentum can be reproduced, as long as we adjust the
non-perturbative parameters.

If we restrict the data to $P^*_{min}=1.5 \gevc$ and use a recent,
independently measured value of $\bar{\Lambda} = 0.35 \pm 0.08 \pm
0.10~\gev$~\cite{Chen:2001fj}, we obtain
   $\lambda_1 = -0.17 \pm 0.06 \pm 0.07 \gev^2$,
a result that is in good agreement with the CLEO value of
$\lambda_1 = -0.236 \pm 0.071 \pm 0.078 \gev^2$. However, if we
take this value of $\lambda_1$ and $\bar{\Lambda} = 0.35 \gev$ and
calculate the moments as a function of $P^*_{min}$, we find a much
smaller momentum dependence than the data indicate (see
Figure~\ref{fig:momentsivar2}). 

In summary, if the assumption is correct that there are significant contributions from charm states with masses extending well beyond the resonance $D^{**}$, the second moment is expected to rise for lower lepton momenta, an effect that is not described by OPE using other independently measured values of the parameter $\bar{\Lambda}$.  On the other hand, we currently do not have adequate knowledge of the branching ratios and mass distributions for higher resonant and mass resonant states in semileptonic B decays.  And their contribution and mass distribution enter critically into the method that has been applied to extract the mass moments.  It is expected that more direct methods to measure moments will reduce this dependency.  Results are expected in the near future. 


Probably the best way to address this problem, is to perform 
more detailed measurements of various
exclusive semileptonic branching fractions for decays to higher mass states.
In addition, we expect to improve the method of determining moments, 
to measure higher mass moments, and to add
moments of the lepton energy spectrum.  Measurements of the
inclusive photon spectrum in $b \to s \gamma$ will also add critical information on
the nonperturbative effects that have impact on the translation of
inclusive decays rates to $|V_{cb}|$, and $|V_{ub}|$.


\begin{thebibliography}{9}

\bibitem{BABARDET} B.~Aubert {\it et al.}, BABAR Collaboration,
Nucl.\ Instrum.\ Meth.\ {\bf A479} (2002) 1.

\bibitem{PEPII}
PEP-II , SLAC-418, LBL-5379 (1993).


\bibitem{ichep2}
B. Aubert {\it et al.}, BABAR Collaboration,
SLAC-PUB-9314, BABAR-CONF-02-029,
Contribution to this Conference (ICHEP 2002),
e-Print Archive: hep-ex/0207084.


\bibitem{PDG00} Review of Particle Properties, Eur. Phys. J. {\bf C15} (2000) 1.


\bibitem{Cronin-Hennessy}
D.~Cronin-Hennessy {\it et al.}, CLEO-Collaboration,
Phys. Rev. Lett. {\bf 87} (2001) 251808.

\bibitem{calvi}
M. Calvi {\it et al.}, DELPHI-Collaboration,
hep-ex/0210046 (2002)

\bibitem{Falk:1998jq}
A.F. Falk and M.E. Luke, Phys. Rev. {\bf D57} (1998) 424., and also private communication.

\bibitem{Chen:2001fj}
S. Chen {\it  et al.}, CLEO Collaboration,
Phys.\ Rev.\ Lett.\  {\bf 87} (2001) 251.

\end{thebibliography}
\end{document}